\documentclass[10pt]{article}

\usepackage{amsmath}
\usepackage{amssymb}

\usepackage{graphicx}

\usepackage{cite}

\usepackage{color} 

\usepackage[labelformat=empty]{caption}

\makeatletter
\renewcommand{\@biblabel}[1]{\quad#1.}
\makeatother

\date{}

\begin{document}

\begin{flushleft}
{\Large
\textbf{Modeling the dynamics of bivalent histone modifications
}
}
\\
Wai Lim Ku$^{1, 2, \ast}$, 
Michelle Girvan$^{1, 2, 3}$, 
Guo-Cheng Yuan$^{4, 5}$,
Francesco Sorrentino$^{6}$,
Edward Ott$^{1, 2, 7}$,
\\
\bf{1} Department of Physics, University of Maryland, College Park, MD, USA
\\
\bf{2} Institute for Research in Electronics and Applied Physics, University of Maryland, College Park, MD, USA
\\
\bf{3} Institute for Physical Sciences and Technology, University of Maryland, College Park, MD, USA
\\
\bf{4} Department of Biostatistics, Harvard School of Publich Health, Boston, MA, USA
\\
\bf{5} Department of Biostatistics and Computational Biology, Dana-Farber Cancer Institute, Boston, MA, USA
\\
\bf{6} Department of Mechanical Engineering, University of New Mexico, Albuquerque, New Mexico 87131, USA
\\
\bf{7} Department of Electrical and Computer Engineering, University of Maryland, College Park, MD, USA
\\
$\ast$ E-mail: Corresponding wlku@umd.edu
\end{flushleft}

\section{Abstract}

Epigenetic modifications to histones may promote either activation or repression of the transcription of nearby genes. Recent experimental studies show that the promoters of many lineage-control genes in stem cells have "bivalent domains" in which the nucleosomes contain both active (H3K4me3) and repressive (H3K27me3) marks. It is generally agreed that bivalent domains play an important role in stem cell differentiation, but the underlying mechanisms remain unclear. Here we formulate a mathematical model to investigate the dynamic properties of histone modification patterns. We then illustrate that our modeling framework can be used to capture key features of experimentally observed combinatorial chromatin states.

\section*{Author Summary}
In a cell nucleus, DNA is wrapped around proteins called histones. These histones can undergo several kinds of epigenetic modifications, either activating or repressing genes that are close to them. In addition, modified histones are believed to induce similar modifications in other nearby histones, leading to complex dynamics. It is thought that regions containing nucleosomes which individually have both modifications that are activating and modifications that are repressing play an important role, e.g., in regulating the activities of developmental genes. Such regions are called bivalent domains. We formulate a mathematical model of the dynamic changes of histone modification patterns. The model is capable of capturing general observed features of bivalent domains, and also yields dynamical evolutions that lead to these features.  For example, we observe that bivalent domains can form through a process of front propagation along a nucleosome array. We anticipate that our model, along with the illustrative results that we have obtained using, it will contribute to the understanding of the dynamical evolutions of combinatorial chromatin states thereby providing insights into the regulatory mechanisms for cell-fate transitions.

\section{Introduction}

Histones can undergo various types of covalent modifications, such as methylation and acetylation, which serve as an additional layer of transcriptional control by mediating the chromatin accessibility and by recruiting regulatory proteins \cite{kou1, Allis}. Experimental studies using chromatin immunoprecipitation followed by massively parallel sequencing (ChIP-seq) have suggested that different cell types can be characterized by different histone modification patterns \cite{Mikkelsen}. 
\\
\\
The molecular mechanisms underlying chromatin state establishment, maintenance, and heritability remain incompletely understood. A number of mechanisms are implicated\cite{Moazed}, including (1) sequence-specific recruitment through interactions between chromatin regulators and DNA binding factors; (2) recruitment of chromatin regulators to existing histone marks; (3) histone marks deposited by transcriptional machineries; (4) RNA mediated  recruitment; and (5) stochasticity associated with DNA replication. However, any single mechanism alone is insufficient for chromatin state establishment \cite{Moazed,Kouzarides2007}.
\\
\\
One of the best characterized chromatin states is a bivalent domain, a segment of the nucleosome array, in which H3K4me3 (an active mark) and H3K27me3 (a repressive mark) coexist on most individual nucleosomes within the domain\cite{bradley_bivalent}. Bivalent domains are thought to be an important feature of stem cells. For example, bivalent domains have been discovered in the promoters of most lineage-control genes in embryonic stem cells, and most of these domains become monovalent upon cell differentiation \cite{Mikkelsen,bradley_bivalent,Zhao2007,Pan2007,Vastenhouw}. Also, a recent study observed that gene activation in the differentiation process occurs in conjunction with the decay of repressive marks in bivalent domains \cite{Chakravarthy}. In particular, one prominent proposal \cite{bradley_bivalent} for the function of bivalent domains is that the H3K27me3 marks act to repress the lineage-control gene in stem cells, while the H3K4me3 marks poise these genes for activation upon cell differentiation. Thus this proposal suggests that activation of these genes in differentiated cells is determined by the existence of bivalent domains in stem cells. These findings indicate the importance of bivalent domains and motivate further study in order to illuminate the underlying principles and mechanisms involved in their formation and evolution.
\\
\\
It has been proposed that the formation of chromatin domains is consistent with a model that includes not only the chemical interactions between histone marks, but also nucleation sites where domains are more likely to form \cite{Hodges}. The dynamics of histone modifications have been studied both theoretically and experimentally for some time \cite{Hodges,Dodd,DavidRus,Mohammad,Ringrose}. In general, histone methylation marks are catalyzed by a variety of methyltransferase enzymes which may act singly or cooperatively. For example, H3K27me3 marks are catalyzed by Ezh2, a core member of the Polycomb group proteins. In addition to the normal stochastic conversion which would be expected from each of these individual enzymes, there is also a feedback process between the histone marks and the enzymes \cite{manchingku}. Existing H3K27me3 marks may attract Polycomb group complexes, which enhance nearby methylation \cite{Margueron2009,Hansen2008}.  A similar recruitment mechanism has also been suggested for H3K4me3 via Trithorax protein complexes (TrxG) \cite{Orlando}. In addition, there exists experimental evidence supporting a negative feedback mechanism between H3K4me3 and H3K27me3 marks via the action of histone demethylases\cite{Welstead,Kooistra,Pasini,Cloos,Kimwam}.
\\
\\
Certain specific DNA sequences may serve as the docking sites of for modification enzymes and may therefore be associated with enhanced local attraction of histone marks\cite{Moazed,gcyuan}. We refer to these as nucleation sites. For example, CpG islands are strongly enriched in bivalent domains in human and mouse embryonic stem cells \cite{Orlando}, and appear to be required for Polycomb binding in certain cases \cite{Mendenhall}.
\\
\\
Recently, \emph{in silico} methods have provided important additional insights for chromatin state inheritance.
Major contributions have been made by Dodd el al. \cite{Dodd} and Sedighi and Sengupta \cite{sengupta}. These paper considered 1-dimensional lattice models in which nucleosomes are allowed to have active or repressive modifications that evolve stochastically and by recruitment. They found that a bistable state with either mostly active nucleosomes or mostly repressive nucleosomes can appear and be heritable, consistent with experimental observations. Subsequently, Hodges and Crabtree \cite{Hodges} found that adding a nucleation site into a model of the above type produces a bounded chromatin domain. Also, in a more recent paper, Binder et al.\cite{rohlf} proposed a model describing binding of catalytic enzymes to DNA and their interaction with histone marks with one aim being explaining length distributions of modiﬁed chromatin regions. These past studies are limited to a single type of histone mark on a nucleosome, whereas it is well-known that gene regulation is governed by combinatorial patterns of multiple histone marks \cite{Sneppen, Allis}. In this paper, we extend previous studies by presenting an approach to model the dynamics of combinatorial chromatin states. This is achieved by allowing each individual nucleosome to carry both active and repressive marks simultaneously. 
\\
\\
In the next section we describe our model. Then, in the Results section, we apply this model to investigate the dynamics of histone modification patterns with the focus on bivalent domains. Discussion and Conclusions are given at the end of the paper.


\section{Methods}

\emph{General framework of our model.} We consider a 1D lattice of $N$ nucleosomes, where there is a nucleosome at each lattice site $i=1,2,...,N$. An actual nucleosome consists of 8 histone protein moleclues, that can be regarded as two identical groups of four each. In what follows we only consider the state of one of these four histone group memebers, namely the H3 histone, which is specifically related to bivalency. Thus, in our model, we represent the state of a nucleosome as being determined by the states of its two H3 histone copies. There are two modification sites in each H3 histone, one which may have an active mark (such as H3K4me3) and the other which may have a repressive mark (such as H3K27me3). As shown in the Supplementary material, this, together with the restriction obtained from experiment \cite{Voigt} that active and repressive marks do not occur simultaneously on the same H3 histone, leads to the six physically distinct nucleosome states depicted in Fig. 1, where we assign the symbols $UU$, $AA$, $RR$, $AU$, $UR$, and $AR$ to the six possible states. In Fig. 1 the circle represents a nucleosome and the vertical ellipses represent H3 histones. From now on, when we say `histone' it is to be understood that we mean an H3 histone. We note that the state $AR$ will play a prominant role in subsequent considerations in Section 4, and we will call a nucleosome in this state a `bivalent nucleosome'.
\\
\\
We then allow each nucleosome state to evolve according to a discrete time ($t$) model, in which from time $t$ to time $t+1$, a nucleosome state changes from state $\sigma$ to state $\sigma^{'}$ with probability $\pi_{\sigma \sigma^{'}}$ . Since the time step $t\rightarrow t+1$ is regarded as small, we assume that, at most, only one modification site may change on each time step. Thus, there are 12 possible transitions among the 6 distinct states (see Fig. 2 which shows the possible transitions). 
\\
\\
\emph{Reduced model} The general framework above can lead to a relatively complex class of models and has many parameters. Thus, for the simulations that we report in this paper, we have adopted the somewhat modest goal of illustrating different types of dynamics that can arise when different nucleosome states interact and compete. With this goal in mind, we now seek an illustrative, but still somewhat plausible, reduction of our general 6-state model. Our reduction is based on the assumption, motivated in the Supplement, that the occurence of nucleosome states having either active marks on both histones ($AA$ in Fig. 1) or repressive marks on both histones ($RR$ in Fig. 1) are unlikely. Thus we consider the idealized case where $AA$ and $RR$ states do not occur. The reduced model then contains only 4 nucleosome states, namely $AU$, $UR$, $AR$ and $UU$ (see Fig. 3A). Referring to Fig. 1, we see that the four states have the following meanings.

\begin{description}
  \item[$AU$:] One histone has an active mark and the nucleosome's other three sites are unmodified.
  \item[$UR$:] One histone has a repressive mark and the nucleosome's other three sites are unmodified.
  \item[$AR$:] One histone has an active modification, while its other site is unmodified. The other histone has a repressive modification, while its other site is unmodified.
  \item[$UU$:] All four sites of the nucleosome are unmodified.
\end{description}

\noindent
\emph{Model Dynamics.} During a cell cycle, we consider the time $t$ states of modeled nucleosomes on our one dimensional lattice and update these states to new states at time $t+1$ through two probabilistic processes that we call ``recruitment conversion'' and ``exhange conversion''. At the conclusion of a cell cycle, ``replication'' occurs, following which a new cycle begins.

\begin{itemize}
\item \emph{Recruitment.} This refers to the recruitment of histone marks to a nucleosome through interaction with neighboring nucleosomes. Recruitment at a site $i$ depends on the states of the nucleosomes in an interval of length $2l$ centered at $i$, and we refer to $l$ as the range of recruitment. We define $f_{X}^{i}$ as the fraction of nucleosomes in this interval which carry a type-$X$ histone mark, where the subscript $f_{x}^{i}$ is $X=A, R$.
If $l\leqslant i\leqslant N-l$, then the recruitment range will span $2l+1$ nucleosomes on our lattice. However, if $i$ is too close to the beginning or the end of the lattice (i.e., $1 \leqslant i<l$ or $N-l< i\leqslant N$, respectively), then the recruitment range will include `phantom' sites $j$ not on the lattice ($j< 1 $ and $j>N$, respectively), and for the purpose of determining $f_{X}^{i}$, we consider such phantom sites $j$ to be in the $UU$ state. The probability of recruitment conversion from $U$ to $X$ at site $i$ is taken to be given by $f_{X}^{i}r_{_{UX}}$, where $r_{_{UX}}$ is a constant describing the strength of the recruitment interaction. On the other hand, the probablity of recruitment conversion from $X$ to $U$ (i.e., mark removal) depends on the concentration of histone marks which are opposite (rather than similar) to $X$ (where we regard $A$ and $R$ as opposites). In this case, the conversion probability is taken to be given by $f_{Y}^{i}r_{_{XU}}$, where $Y$ is $R$ if $X$ is $A$ and vice versa (see Fig. 3B). Note that in our model we allow $r_{_{XU}}$ to differ from $r_{_{UX}}$ because different enzymes are recruited for the addition and removal of histone marks.
\item \emph{ Exchange.} Unlike the recruitment process, the exchange process refers to histone modifications which occur spontaneously, independent of the states of nearby nucleosomes. The probabilities for exchange conversion are denoted by $p_{_{UA}}$, $p_{_{UR}}$, $p_{_{AU}}$, and $p_{_{RU}}$ (see Fig. 2C). In particular, we think of $p_{_{AU}}$ and $p_{_{RU}}$ as corresponding to the histone turnover process, and $p_{_{UA}}$ and $p_{_{UR}}$ as corresponding to processes involving nucleation sites (See Table 1).
\item \emph{DNA replication.} When DNA replication occurs, we imagine that in the real situation the parental nucleosomes are randomly assigned to one of the two daughter strands at the same site as that which they occupied on the parental strand, while the corresponding site on the other strand is assigned an unmodified nucleosome (i.e., a nucleosome in the $UU$ state). This scenario is supported by an experimental observation \cite{Radman}. In our model, we do not follow both daughter strands. Rather we follow just one. Thus, with probability 1/2, our model replication process randomly replaces each nucleosome with an unmodified ($UU$) nucleosome. This model DNA replication occurs periodically with a period equal to the `cell cycle time' $\tau$. This is similar to how replication is modeled in \cite{Dodd}. 
\end{itemize}
In accord with the above recruitment and exchange processes, during a cell cycle, our model gives appropriate equations for the probablities $P_{XY}^{i}(t+1)$ that nucleosome $i$ is in state $XY={UU, AU, UR, AR}$ at time $t+1$, given the state of the lattice at time $t$. After the probablities $P_{XY}^{i}(t+1)$ are determined the state ($UU$, $AU$, $UR$ or $AR$) of each nucleosome $i$ is randomly chosen according to the probabilities $P_{XY}^{i}(t+1)$, thus determining the state at time $t+1$. Letting $\delta_{XY}^{i}(t)=1$ if nucleosome $i$ is in state $XY$, and $\delta_{XY}^{i}(t)=0$ if nucleosome $i$ is not in state $XY$, our model equations for the probabilities are

\begin{eqnarray}
P_{AU}^{i}(t+1)&=&2[f_{A}^{i}(t)r_{_{UA}}+p_{_{UA}}^{i}]\delta_{UU}^{i}(t) 
\nonumber + [f_{A}^{i}(t)r_{_{RU}}+p_{_{RU}}]\delta_{AR}^{i}(t) 
\\ \nonumber &+& \{1- [f_{R}^{i}(t)(r_{_{AU}}+r_{_{UR}})+p_{_{AU}}+p_{_{UR}}^{i}]\}\delta_{AU}^{i}(t),
\\ \nonumber
P_{UR}^{i}(t+1)&=&2[f_{R}^{i}(t)r_{_{UR}}+p_{_{UR}}^{i}]\delta_{UU}^{i}(t) 
\nonumber + [f_{R}^{i}(t)r_{_{AU}}+p_{_{AU}}]\delta_{AR}^{i}(t) 
\\ \nonumber &+& \{1- [f_{A}^{i}(t)(r_{_{RU}}+r_{_{UA}})+p_{_{RU}}+p_{_{UA}}^{i}]\}\delta_{UR}^{i}(t),
\\ \nonumber
P_{AR}^{i}(t+1)&=&[f_{R}^{i}(t)r_{_{UR}}+p_{_{UR}}^{i}]\delta_{AU}^{i}(t)
\nonumber + [f_{A}^{i}(t)r_{_{UA}}+p_{_{UA}}^{i}]\delta_{UR}^{i}(t) 
\\ \nonumber &+& \{1-[f_{A}^{i}(t)r_{_{RU}}+f_{R}^{i}(t)r_{_{AU}}+p_{_{RU}}+p_{_{AU}}]\}\delta_{AR}^{i}(t),
\\ \nonumber
P_{UU}^{i}(t+1)&=&1-\{P_{AU}^{i}(t+1) + P_{UR}^{i}(t+1) + P_{AR}^{i}(t+1)\}.
\end{eqnarray}
Consistent with our assumption that at most one site on a nucleosome can change state in one time step, our choice of parameters satisfies $r_{_{XY}}, p_{_{XY}} \ll 1$. Note that $f_{A}^{i}(t)$ and $f_{R}^{i}(t)$ depend on the lattice state in a neighborhood of site $i$ within the range of recruitment specified in the second bullet above.
\\
\\
In subsection 4 of the Results section, where we treat localization of $AR$ states, we allow the exchange transitions probabilities $p_{_{XY}}^{i}$ to vary from site to site, but everywhere else we consider $p_{_{XY}}^{i}$ to be the same at each site, $p_{_{XY}}^{i}=p_{_{XY}}$.
\\
\\
\emph{Simulation Parameters.} To assign roughly reasonable values to the parameters $r_{_{XY}}$ and $p_{_{XY}}$, we first consider that our model time step, $t\rightarrow t+1$, corresponds to a real time step $\Delta t=2$ min. We have numerically verified that our simulation results are independent of our choice of $\Delta t$ so long as $\Delta t$ is sufficiently small. To estimate a rough range for the parameters $r_{_{XY}}$ and $p_{_{XY}}$, we set $p_{_{XY}}$, $r_{_{XY}}\approx (\Delta t/T)$, where $T$ is the characteristic time scale of the relevant process (see Table 1), and, as required, the $\Delta t$ that we have chosen is such that $\Delta t/T$ is small compared to one for all such processes. We fix as many parameters (Table 1) as possible using experimental information (see Table 2). Because the authors are not aware of any experimental measurements of the characteristic time for recruitment demethylation and methylation via exchange, we will consider these probabilities as free parameters in our numerical simulations below.  Previous work \cite{Deal} suggests that the loss of active marks is faster than the loss of repressive marks. In particular, it has been shown that nucleosome turnover is faster in regions bound by trithorax-group proteins. Therefore, we selected the model parameters so that all rates associated with active mark are faster than those associated with the repressive mark. Specifically, we assume that $r_{_{UR}}/r_{_{UA}} = p_{_{UR}}/p_{_{UA}} = r_{_{RU}}/r_{_{AU}} = p_{_{RU}}/p_{_{AU}} = 0.5$ in the simulation (when nonzero). Regarding the cell cycle, for embryonic stem cells the cell cycle length is about 12 hours, which, with our $\Delta t=2$ min, corresponds to $360$ time steps of our discrete time model per cell cycle. Finally, motivated by Ref.\cite{Radman}, we take $l=2$, corresponding to a fairly short range of recruitment.

\section{Results}

We now illustrate the utility of our model by employing it to investigate dynamic changes of histone modification patterns. As described in the Introduction, both nucleation sites and recruitment of methylation may be involved in the establishment of bivalent domains. As noted above, we suggest that certain nucleosomes act as nucleation sites during the early stages of development. These nucleation sites may be instrumental in the formation of bivalent domains. We incorporate nucleation sites into our model by assigning them a higher value of $p_{_{UA}}$ and $p_{_{UR}}$ than other sites, and we model the absence of nucleation sites by lowering its value of $p_{_{UA}}$ and $p_{_{UR}}$.
\\
\\
In Sections 4.1 and 4.2, we discuss the formation and decay of $AR$ states with different initial conditions in the absence of nucleation sites. In Section  4.3, we study the effect of nucleation sites on dynamics of the formation of $AR$ states. Finally, in Section 4.4, we consider how varying the cell-cycle length affects $AR$ states. Taken together, these analyses demonstrate the utility or our model for systematic investigation of the dynamic properties of bivalent domains.

\subsection{Formation of AR states}

The formation of bivalent domains has been experimentally observed in studies of the early stages of embryogenesis \cite{Vastenhouw2010922} and in studies of cell reprogramming \cite{Wernig2007318}. In particular, studies of cell reprogramming observe this formation process to be gradual\cite{Polo}.
\\
\\
In this section we use our model to simulate the formation of regions that are dense with $AR$ states, and we identify such regions with bivalent domains. In the simulations, we take $p_{_{UA}}=p_{_{UR}}=0$ for all nucleosomes and fix $r_{_{UA}}$= 0.046 (corresponding to an H3K4me3 methylation timescale of 30 mins) and $p_{_{AU}}=0.005$. Also, $r_{_{AU}}$ and $r_{_{RU}}$  are considered to be very small (for simplicity, we set $r_{_{AU}}$ = $r_{_{RU}}$ =0), so that the $AR$ states can be established and persist for a long time. For the initial state of the lattice in the simulations, we consider a situation where there are a relatively small number of nucleosomes in $AR$ states, with all other nucleosomes initially in the $UU$ state. In particular, we choose the initial number of $AR$ nucleosomes to be five (out of the 80 nucleosomes on the lattice), and we study how $AR$ states spread to other nucleosomes on the lattice. To investigate the effect of the initial spatial distribution of $AR$ nucleosomes, we consider two extreme cases: \emph{the localized case} in which all five initial $AR$ state nucleosomes are located at five consecutive nucleosome sites in the center of the lattice, and \emph{the delocalized case} in which the five initial $AR$ state nucleosomes are located at equally spaced sites spanning the entire lattice (at sites 1, 20, 40, 60, 80).
\\
\\
Fig. 4 shows results for the space-time evolution of the distribution of nucleosomes for both \emph{localized} (left column of figure panels ) and \emph{delocalized} (right column of figure panels) initial states. Fig. 4C shows space-time plots for the four types of nucleosomes in a typical single run, while Figs. 4A-B show average space-time plots of the level of $AU$ and $AR$ nucleosomes, that is, the fraction of runs for which the nucleosome is in the indicated state. The average level of $UR$ nucleosomes (not plotted) is low everywhere all the time (dark blue, in terms of the color scale of Figs. 4A and B). Note that, in Figs. 4A-B, the regular drops of the levels of the indicated nucleosomes every 360 time steps (corresponding to the start of a new cell cycle) are due to the inserted of $UU$ nucleosomes in the DNA replication process. In Fig. 4A, for the localized case (corresponding to the left panel figure), the $AR$ nucleosomes spread over the lattice via a propagating front \cite{sengupta} manifested by the approximately straight lines of the color transition boundaries eminating from the space-time point at the center of the lattice at time  $t=0$. For the delocalized case (right panel of Fig. 4A), $AR$ nucleosomes spread over the lattice via individual propagating fronts emanating from the five initial $AR$ sites. These fronts merge near the end of the first cell-cycle (time $\approx$ 300), but the system takes longer time (time $\approx$ 1250) to reach a final equilibrium distribution. The model results show that, while the space time evolution of the distribution of $AR$ nucleosomes is dependent upon the initial condition, the time it takes to establish a final equilibrium distribution is comparable and relatively long for both the localized and delocalized cases. This may have relavance to the experimental observation of Ref. \cite{Polo} that the establishment of bivalent domains is gradual. 
\\
\\
For the localized case, there appears to be two fronts, a fast $UU\rightarrow AU$ front (corresponding to the blue to yellow transition in the left panel of Fig. 4C), followed by a $AU\rightarrow AR$ front (yellow to red transistion in the left panel of Fig. 4C) that propagates at a slower speed than the $UU\rightarrow AU$ front. The slow $AU\rightarrow AR$ front is clearly seen in the left panels of Figs. 4A and B, while the $UU\rightarrow AU$ front is evident in the left panel of Fig. 4B. These two fronts propagate symmetrically in space in the average space-time plots (Fig. 4A and 4B) but, due to fluctuations, more asymmetrically in space in the single run plot (see left panel of Fig. 4C). Examining a range of parameters, we find that the fastest front corresoponds to either a $UU\rightarrow AU$ transition (as in Fig. 4B) or a $UU\rightarrow RU$ transition (not shown). For the delocalized case, we also observe that the spreading of the active marks is faster than that of the repressive marks. This can be easily seen from the typical single run plot in the right panel of Fig. 4C.
\\
\\
Finally, we also studied the effects of varying the number of $AR$ nucleosomes in the initial condition on the above simulations. Using the same parameters values as above, we plot (Fig. 5) the final average fraction of $AR$ nucleosome at the end of the final simulated cell cycle (~10 cell cycles) as function of the initial number $m$ of $AR$ nucleosomes which are taken to occupy the $m$ nucleosome sites in the center of the lattice. As shown in Fig. 5, the average fraction of final $AR$ nucleosomes initially increases with increasing $m$. We observe that past  $m \ge 4$ the value is essentially constant up to $m=80$ with $AR$ nucleosomes spanning the whole lattice. For a given $m$, each simulation can be categorized into two groups, (1) the final spatial average level of $AR$ nucleosomes is approximately equal to the corresponding large $m$ limiting value, or (2) all $AR$ nucleosomes vanish. Thus at low $m$, the value plotted on the vertical axis of Fig. 5 can be thought of as the limiting larger-$m$ value (basically the value  at $m=4$) multiplied by the fraction of runs in category (1). In the early stage of a simulation, the spreading of histone marks compete with the loss of histone marks via histone turnover. If either type of mark is lost totally, it cannot recover (i.e., the run is in catagory 2). On the other hand, we find that histone marks do not die out if there are enough of them on the lattice (the run is then in catagory 1). As a result, the average fraction of $AR$ nucleosomes is larger with larger $m$, and with smaller $p_{_{AU}}$ and $p_{_{RU}}$ (compare the red and blue plots in Fig. 5). The above simulations suggest that in order for $AR$ states to form when $p_{_{UA}}$ and $p_{_{UR}}$ are small, a sufficient number of initial $AR$ nucleosomes is required.

\subsection{Decay of AR states}

In this section we use our model to simulate the decay of $AR$ states. All parameters are the same as in section 4.1 except that $r_{_{AU}}$  and $r_{_{RU}}$ are taken to be non-zero. This is motivated by experimental findings that recruitment of demethylases is important for the decay of bivalent domains \cite{Pasini,Cloos}, and occurs during cell differentiation. Also, we consider an initial condition in which all nucleosomes are in $AR$ states. Results are shown in Figs. 6-8 for different values of $r_{_{AU}}$ and $r_{_{RU}}$ keeping their ratio fixed at $r_{_{AU}}/r_{_{RU}}=2$.
\\
\\
Fig. 6 shows results for the space-time evolution of the distribution of all four nucleosome states $AR$, $UR$, $AU$, and $UU$ for three values of $r_{_{AU}}\equiv 2r_{_{RU}}$. In Fig. 6A, for the case $r_{_{AU}}=0.004$, the initial level of $AR$ nucleosomes rapidly (in about one cell-cycle) drops to a lower level of $AR$ nucleosomes, but there still remains a substantial presence of $AR$ nucleosomes which persists to the end of the run. In contrast, for both $r_{_{AU}}$= 0.016 and $r_{_{AU}}$= 0.034, where there is again similar very rapid decreases of the level of $AR$ nucleosomes, now the final level is essentially zero. In addition, it is seen that the level of $AR$ nucleosomes takes longer to fully decay for $r_{_{AU}}=0.016$ than for $r_{_{AU}}=0.034$. The latter case is consistent with the experimental observations \cite{Chakravarthy,Polo} that an essentially complete loss of bivalent domain can occur very rapidly. To further explore how the decay of $AR$ states depends on the recruitment demethylation rates, we plot the fraction of simulation runs that have at least one $AR$ nucleosome on the lattice as a function of time in Fig. 7, and the the final average fraction of $AR$ nucleosomes (averaged over 1000 runs) as a function of $r_{_{AU}}$ in Fig. 8. Comparing Fig. 6A to Fig. 7, we observe that the fraction of runs with at least one $AR$ nucleosome plotted in Fig. 7 shows a slower decay compared to the decay of $AR$ levels in Fig. 6A. This suggests that lineage-control genes in bivalent domains may become active without the full destruction of repressive marks. In Fig. 8, as might be anticipated, we observe that, in general, smaller histone turnover ($p_{_{AU}}$) and smaller recruitment demethylation rate give a higher final average fraction of $AR$ nucleosomes. Also, the value of $r_{_{AU}}$ at which the average fraction of $AR$ nucleosomes drops to zero is lower for larger $p_{_{AU}}$. Our results suggest that a large recruitment demethylation rate in a cell is important for cell differentiation. This is consistent with experimental findings \cite{Pasini,Cloos}. 
\\
\\
In a real situation, a change from low to high values of the recruitment demethylation rates during cell differentiation will take place by processes not included in our model, and these processes may take some time. Thus our simulation use of constant non-zero initial $r_{_{AU}}$ and $r_{_{RU}}$ results in a determination of the characteristic decay time associated only with processes that are included in our model, and the true decay rate of $AR$ state nucleosomes may be longer than this time due to the finite time for $r_{_{AU}}$ and $r_{_{RU}}$ to change. Overall, we observe that the decay determined from our model of $AR$ state nucleosomes in response to high initial value of recruitment demethylation rate is relatively fast, as compared to the time that it takes to establish $AR$ states spanning the lattice in Section 4.1. We conclude from this that processes included in our model do not prevent rapid decay of $AR$ state nucleosomes, and that rapid decay, as seen in experiments \cite{Chakravarthy}, can occur in response to rapid increase of $r_{_{AU}}$ and $r_{_{RU}}$.
\\
\\
In addition, it is interesting to emphasize the probabilistic nature of these results. For example, Fig. 6D shows results of typical single realizations. This figure also shows that the final state for $r_{_{AU}}=0.016$ is different from that for  $r_{_{AU}}$= 0.034. For the case $r_{_{AU}}$= 0.016, we observe that $AU$ nucleosomes are dominant in the lattice at the end of the simulation (see also the second panels of Figs. 6B and 6C). However, for the case of  $r_{_{AU}}$= 0.034 at long time, green regions of $UR$ nucleosomes form at the upper edge (see third panel of Fig. 6D), while the $AU$ nucleosomes are at the lower edges. This is because the $AU$ and $UR$ states can both be stable for this combination of parameters (see third panels of Figs. 6B and 6C). Our results suggest that the strength of the recruitment demethylation (i.e., the values of $r_{_{AU}}$ and $r_{_{RU}}$) is not only important for the decay of bivalent domains, but also strongly influences the possible final state following decay.

\subsection{The localization of $AR$ states}

The next issue that we discuss is the effect of nucleation sites (i.e., in our model, $p_{_{UA}}$ $=$ $p_{_{UR}}>0$ at these sites). The existence of such sites is suggested by the finding \cite{Moazed,gcyuan} that DNA specific sequences can recruit protein binding factors like TF which in turn recruit histone marks to the DNA. In section 4.1, we took $p_{_{UA}}=p_{_{UR}}=0$, and we found that $AR$ nucleosomes either span the whole lattice or disappear. Although similar broad bivalent domains are observed, narrow bivalent domains are also detected in some experiments \cite{Kimwam,Mikkelsen}. Although a recent model \cite{Hodges} has previously been used to simulate the dynamics of localized histone modification domains, that model allowed only a single type of histone modification, and therefore it cannot address the dynamics of bivalent domains. Using our model, we will be able to analyze interactions among the placements of active and repressive histone marks, histone turnover rate, and crosstalk between active and repressive histone marks. We consider $p_{_{UA}}$ and $p_{_{UR}}>0$ for the central nucleosome (corresponding to the case that the central nucleosome is a nucleation site), and, using our previous parameter ratios (i.e., $r_{_{UR}}/r_{_{UA}} = p_{_{UR}}/p_{_{UA}} = r_{_{RU}}/r_{_{AU}} = p_{_{RU}}/p_{_{AU}} = 0.5$), we explore the parameter space regions for which our model reproduces narrow and broad distributions of $AR$ nucleosomes.
\\
\\
We consider cases of both relatively small and relatively large recruitment demethylation rates ($r_{_{AU}}$ and $r_{_{RU}}$). The former and latter choices are meant to simulate cell environments far before, and during, cell differentiation, respectively.  For the case of small recruitment demethylation rate, Figs. 9A-B show plots of the fraction of $AR$ nucleosomes averaged over 2000 simulations. Figs. 9A-B demonstrate narrow (left panels of Figs. 9A and 9B) and broad (right panels of Figs. 9A and 9B) distributions of $AR$ nucleosomes. The widths of these bounded distributions reflect the balance between the continuous placement of histone marks on the nucleation site, the spreading of histone marks by the recruitment process, and the destruction of histone marks via exchange \cite{Hodges}. From the simulations, we find that the width of the distributions of $AR$ nucleosomes depends more on $p_{_{AU}}$ and $p_{_{RU}}$, which they are inversly related to the width of the $AR$ distribution. On the other hand, the amplitude of the distributions depends more on $p_{_{UA}}$ and $p_{_{UR}}$ (i.e., the continuous placements of histone marks on the center nucleosome) (Figs. 9A-B).
\\
\\
Next, we did simulations using the same parameters as in Fig. 9A but with larger recruitment demethylation rates ($r_{_{AU}}$ and $r_{_{RU}}$). The results are shown in Fig. 9C. Both of the corresponding distributions in Fig. 9A become narrower in Fig. 9C. In particular, the changes in the broad distribution (right panel) is particularly dramatic. This suggests that it may be easier to see changes in the broad bivalent domain than the narrow one during cell differentiation in experiments. Overall, our results demonstrate that nucleation sites can be responsible for the onset of bounded domains of $AR$ nuclesomes. Also, narrow distributions can be obtained via either enhanced histone demethylation via exchange or via enhanced recruitment.

\subsection{The effects of cell-cycle length on the stability of $AR$ states}

During DNA replication, the nucleosomes, along with their associated histone marks, must be dissociated from the mother strand. How these marks are reassembled to the newly synthesized strands remains poorly understood. Recent studies suggest that the nucleosome, along with their associated marks, are randomly distributed to daughter strands \cite{Radman}. In this section, we use our model to study the impact of DNA replication on the level of $AR$ nucleosomes.
\\
\\
We choose parameters which correspond to cell environments during the formation of bivalent domains (see Fig. 10). Also, we assume that nucleation sites lose their properties at the very beginning of the simulations, so that there are no nucleation sites. We then vary the cell cycle lengths from  6 hours to 24 hours, which corresponds to varying the cell cycle length from that in stem cell to that in differentiated cells. We run the simulations for 10 cell cycles such that the average level of $AR$ nucleosomes over a cell cycle reaches a stable value. Fig. 10 shows the average level of $AR$ nuclesomes as a function of cell cycle length, where these levels are computed by averaging the number of $AR$ nucleosomes at the end of the simulations over the lattice and over all simulation runs. Fig. 10 shows that the average level of $AR$ nucleosomes is, in general, larger for longer cell-cycle. This result is expected, since there is more time for the lattice to recover from the loss of $AR$ nucleosomes, caused by DNA replication, when the cell-cycle is longer. This result is consistent with the experimental finding that higher levels of histone marking are observed when the length of the cell cycle increases \cite{Calder13}.
\section{Discussion}

Development of computational models of bivalent domain dynamics can help to elucidate the mechanism of chromatin domain formation, and give insight for formulating and analyzing experimental studies. In this paper we introduce a model that incorporates multiple histone marks on a nucleosome and the interactions among these marks. We have illustrated the potential use of our model by employing it to investiate the dynamics of bivalent domains, with the following results.
\\
\\
In Section 4.1 we discussed the formation of bivalent domains, which are modeled as $AR$ states. In our model, we regard the existence of $AR$ nucleosomes as analogous to the existence of bivalent domains. We simulate the formation of $AR$ nucleosomes by considering small recruitment demthylation such that the existence of $AR$ nucleosomes can persist for very long time. We find that the $AR$ nucleosomes can be spread over the lattice via  propagating fronts. Also, we find that a minimum number of initial $AR$ nucleosomes is required to guarantee the formation of $AR$ states if $p_{_{UA}}=p_{_{UR}}=0$. On the other hand, in Section 4.2, we discussed the decay of $AR$ states. We consider all nucleosomes are in $AR$ state initially, and we find that $AR$ nucleosomes decay more rapidly as the recruitment demethylation rates are increased. We find that our model allows rapid decay of $AR$ nucleosomes, as seen in experiments (i.e., bivalent domains decay in about 24 hours)\cite{Chakravarthy}.  
\\
\\
It has been observed that bivalent domains are mostly localized in specific DNA regions, and several studies have shown that the width of bivalent domains can vary between different genome locations \cite{manchingku,Orlando,Kimwam}. In particular, a recent study found that, bivalent domains could generally be classified as being either narrow or broad \cite{Kimwam}. To simulate how such situations can arise, in Section 4.3, we take the central nucleosome to be a nucleation site where $p_{_{UA}}$ and $p_{_{UR}}$ are non-zero. We demonstrate that demethylation, either via exchange or recruitment can decrease the width of the resulting $AR$ region. 
\\
\\
Finally, in Section 4.4, we examined how the cell-cycle length affects the level of $AR$ nucleosomes. Consistent with experiental observations \cite{Calder13}, our simulation results suggest that the level of $AR$ nucleosomes may be higher when the cell cycle is longer. 
\\
\\
Although the real dynamics of bivalent domains is complex, our simple dynamical model offers an opportunity to test and compare hypotheses that may motivate future experiments. As more accurate data become avalible, we anticipate that parameters in our model will be better determined, thus enhancing the model's predictive capability. We hope that our work modeling bivalent domains will help to further understanding of the underlying principles of bivalent domain dynamics, important in both cell reprogramming and cancer biology.

\section{Supplemental}

\subsection{Model}

\emph{6-state model.} Refer to Fig. S1. Circles in the figure represent nucleosomes. A nucleosome contains two histone copies represented by the vertically oriented ellipses. Each histone has a site (represented by the upper half of the ellipse) that can be either unmodified (symbolized by $u$) or have an active mark (symbolized by $\alpha$) and another site (represented by the lower half of the ellipse) that can be either unmodified (symbolized by $u$) or have a repressive mark (symbolized by $\rho$). Each of the four modification sites in a nucleosome can be in one of two states (modified or unmodified), yielding the $2^{4}=16$ possibilities that are shown in the figure panels, (a)-(p). Panels grouped together by the curly brakets in the figure represent the same physical nucleosome state, e.g., panels (e) and (f) are considered to represent the same physical nucleosome state since (f) results from (e) by interchange of the left and right histone ellipses. There are six such pairs. Thus there are 10 physically distinct nucleosome states. In addition, experiments indicate that active and repressive marks do not occur simultaneously on the same histone \cite{Voigt} (i.e., $\alpha$ and $\rho$ do not occur in the same ellipse). This eliminates the possibilities depicted in panels (d) and (k)-(p). Thus we arrive at 6 possible states which we label $UU$, $AA$, $RR$, $AU$, $UR$, and $AR$ as shown in the figure.
\\
\\
\emph{Reduced model.} We now introduce a reduction of the above 6-state model to a more simple model. Our reduction is motivated by a limited number of simulations of the 6-state model in which we found that the experimentally observed bivalent nucleosome state ( $AR$) tended to be absent unless the $AA$ and/or $RR$ states were suppressed (i.e., $\pi_{\sigma\sigma^{'}}$ is low for the transition $\sigma=AU\rightarrow \sigma^{'}=AA$ and the transition $\sigma =UR \rightarrow \sigma^{'}=RR$). This is consistent with a recent experimentally motivated hypothesis that the existence of the asymmetrically modified nucleosome states, $AU$ and $UR$, are important for the formation of bivalent domains\cite{Voigt}.
\\
\\
One way of understanding this is to note from Fig. 2 that the $AA$ state competes with the $AR$ state for conversion from the $AU$ state, and the $RR$  state similarly competes with the $AR$ state for conversion from the $UR$  state. This suggests that if we want to allow for the occurence of the experimentally observed $AR$  state, we could chose parameters in our six state model such that the transition rate from $AU$ to $AA$  is sufficiently smaller than the transition rate to $AR$. Similarly we would want the transition rate from $UR$ to  $RR$ to be sufficiently smaller than the transition rate to $AR$. Thus, to make the model more tractable, we employ a further simplification and consider the idealized case in which $AA$ and $RR$  are completely suppressed. That is, in terms of our 6-state model, we set $\pi_{\sigma\sigma^{'}}=0$ for the transition $\sigma=AU\rightarrow \sigma^{'}=AA$ and the transition $\sigma =UR \rightarrow \sigma^{'}=RR$. In this formulation, $AA$ and $RR$ states do not occur, and the 6-state model reduces to a 4-state model. 

\subsection{Another example of localization of $AR$ states related to our results in Section 4.3}

We note that our result in Fig. 9 is not consistent with experiment in that in Fig. 9 the active marks are more extensive than the repressive marks, while Ref. \cite{bradley_bivalent} shows that the opposite situation holds in experiment. We note, however, that, as shown in Fig. S2, for other reasonable parameter choices, we can also obtain states for which the repressive marks are more extensive than the active marks (consistent with \cite{bradley_bivalent}).


\section*{Acknowledgments}
We thank Nicole Francis and Shane Squires for discussions.

\bibliography{biref}


\begin{figure}[!ht]
\begin{center}
\includegraphics[width=3in]{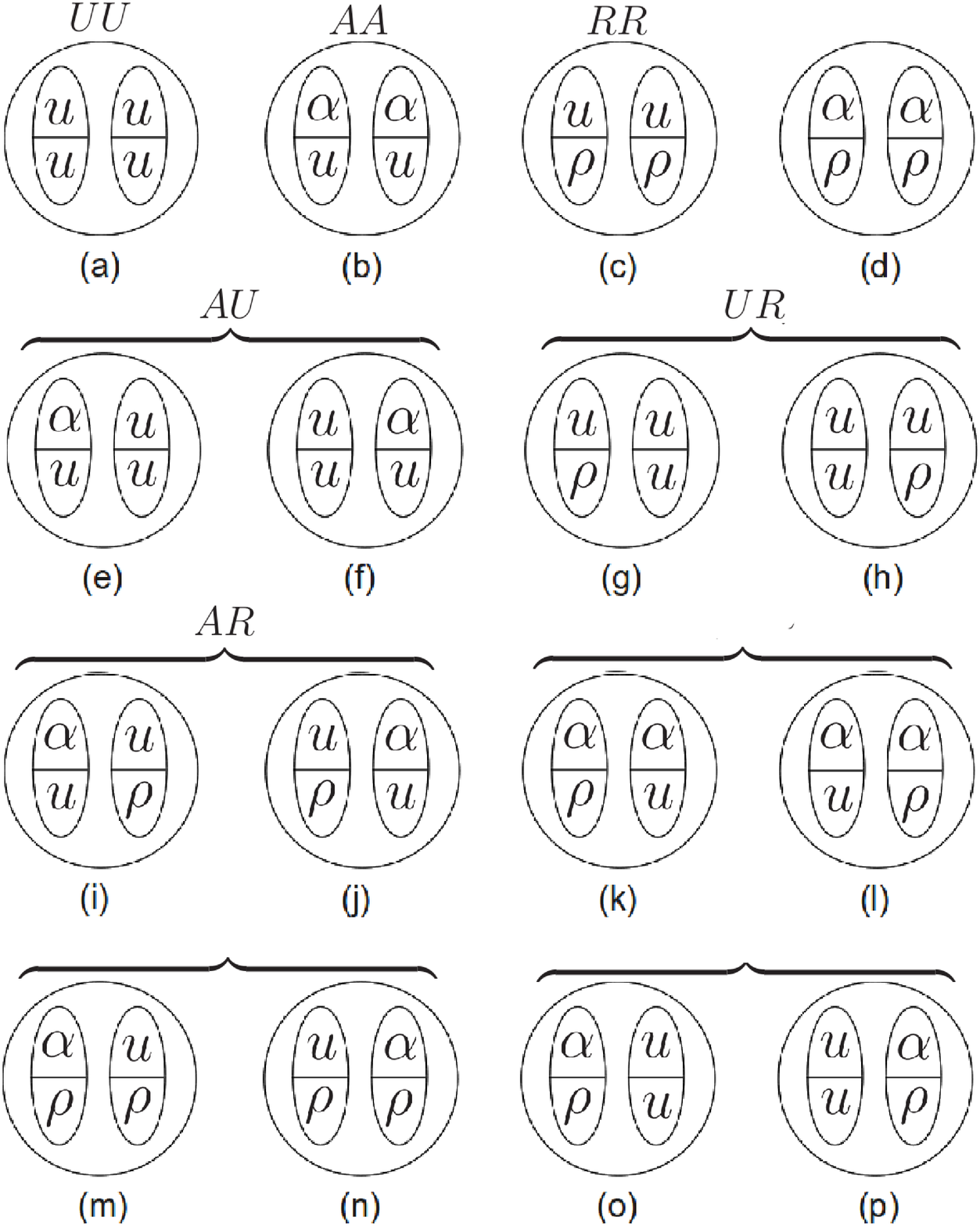}
\end{center}
\caption{
{\bf Figure S1.} Illustration for the explanation of the states of the 6-state model.
}
\label{figureS1}
\end{figure}

\begin{figure}[!ht]
\begin{center}
\includegraphics[width=4in]{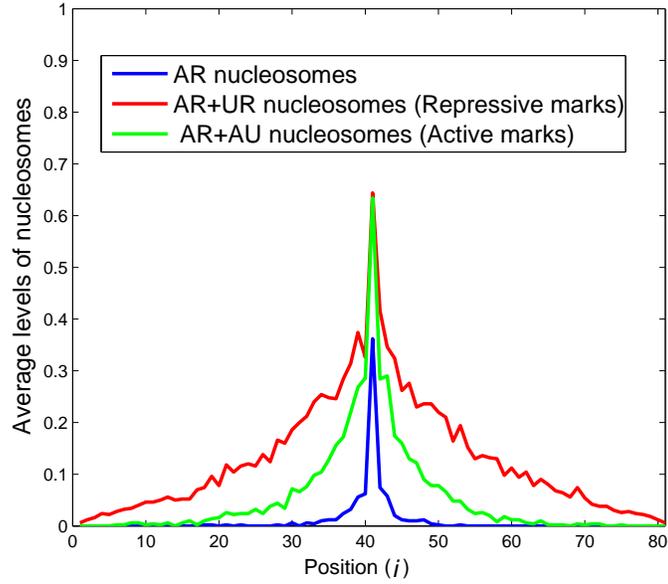}
\end{center}
\caption{
{\bf Figure S2.} This plot illustrates that the 4-state model described in the main text can simulate bivalent domains (blue) in which the active mark (green) is less extensive than the repressive mark (red) (i.e., the bivalent domains (blue) are buried in the repressive domains (red)). The details of simulation can be referred to Section 4.4 in the main text. Here, distributions of $AR$ nucleosomes (blue), $AR+UR$ nucleosomes (red), and $AR+AU$ nucleosomes (green) are plotted at the end of the simulation runs (time =5000). The average levels of nucleosomes are averaged over 1000 simulation runs. In the simulation, $p_{UA}^{i=40}=0.03$ and $p_{UR}^{i=40}=0.015$. The other parameters are  $r_{UA}=0.029$, $r_{UR}=0.021$, $p_{AU}=0.025$, $p_{RU}=0.015$, $r_{AU}=0.004 $ and $r_{RU}=0.002$. 
}
\label{supplefigure1}
\end{figure}

\begin{figure}[!ht]
\begin{center}
\includegraphics[width=3in]{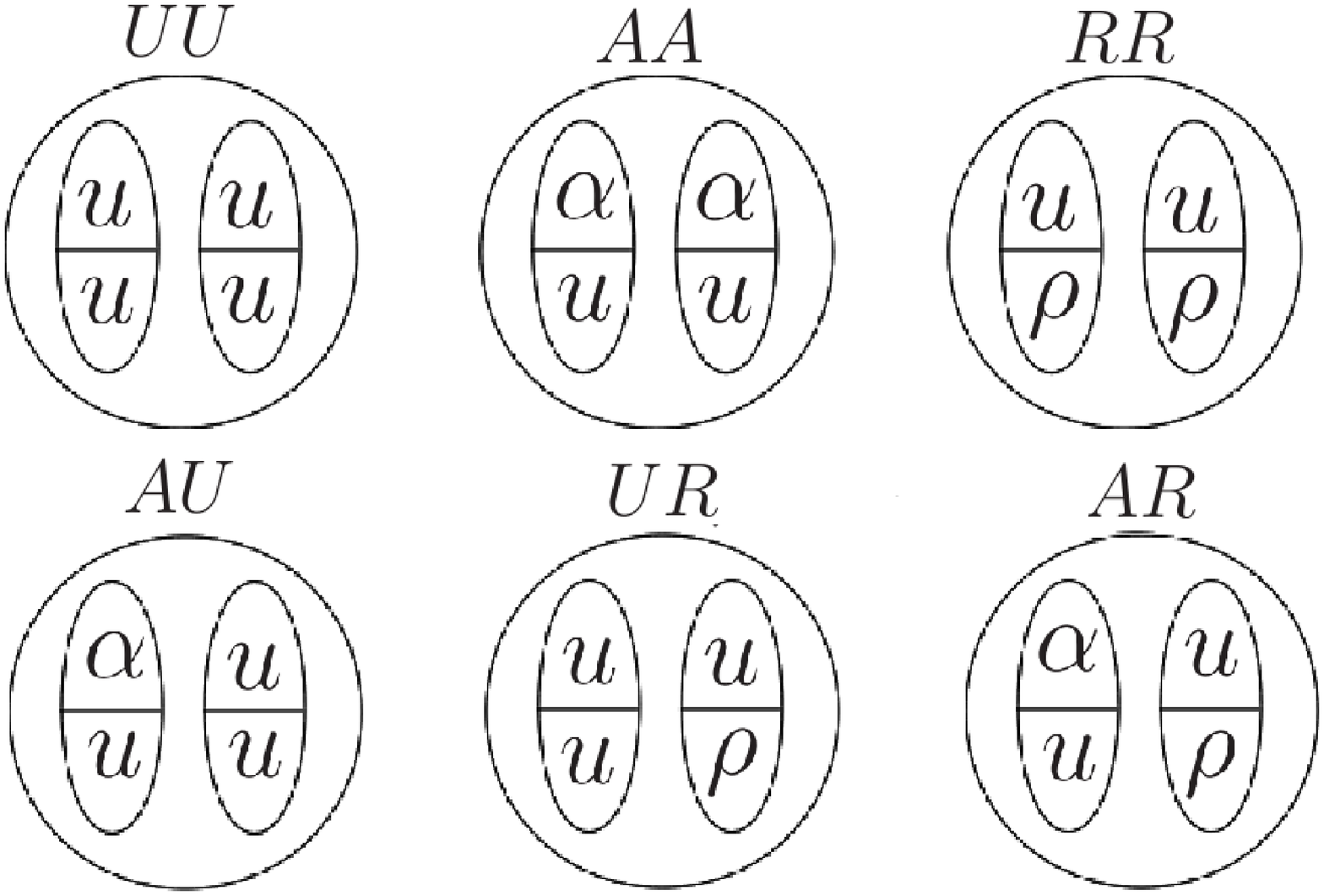}
\end{center}
\caption{
{\bf Figure 1.} Illustration of the states in the 6-state model. Circles represent nucleosomes. A nucleosome contains two histones copies represented by the vertically oriented ellipses. Each histone has a site (represented by the upper half of the ellipse) that can be either unmodified (symbolized by $u$) or have an active mark (symbolized by $\alpha$) and another site (represented by the lower half of the ellipse) that can be either unmodified (symbolized by $u$) or have a repressive mark (symbolized by $\rho$). (Note that the physical nucleosome states labled $AU$, $UR$ and $AR$ could be just as well depicted by interchangeing the left and right ellipses within the respective circles.)
}
\label{figure1}
\end{figure}


\begin{figure}[!ht]
\begin{center}
\includegraphics[width=4in]{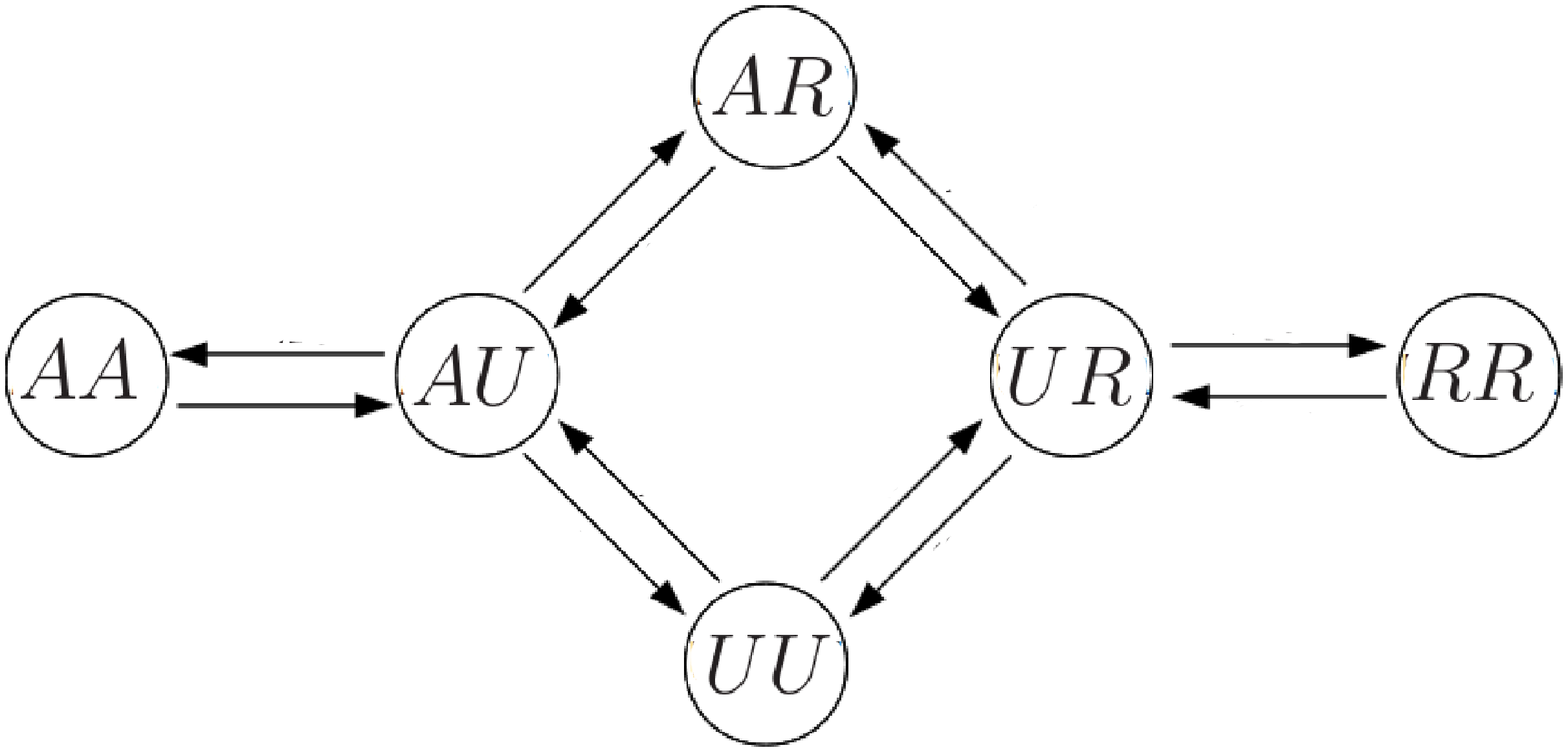}
\end{center}
\caption{
{\bf Figure 2.} Transitions among the 6 distinct states in the 6-state model are indicated by arrows. The time step is supposed to be chosen small enough that only one site of the four nucleosome modification sites shown in Fig. 1 may change on each time step.
}
\label{figure2}
\end{figure}

\begin{figure}[!ht]
\begin{center}
\includegraphics[width=5in]{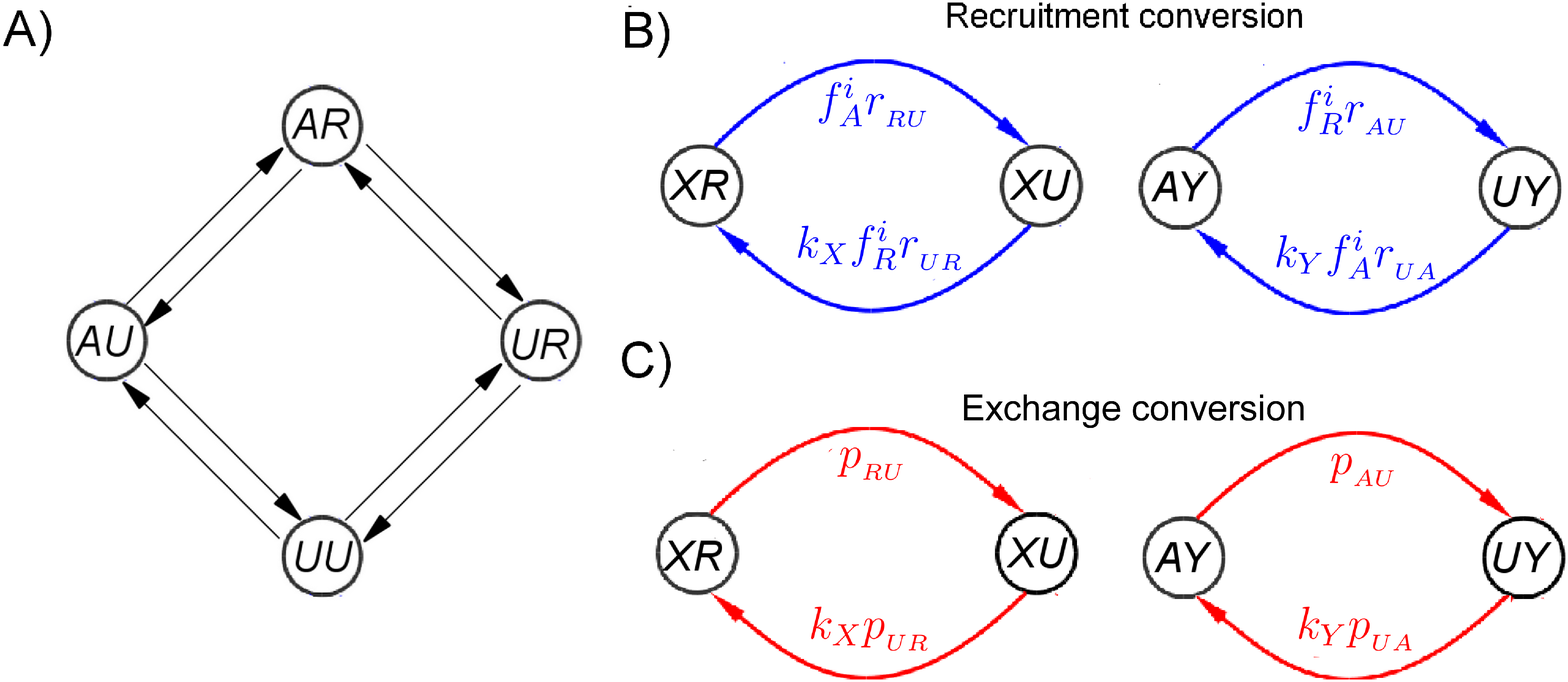}
\end{center}
\caption{
{\bf Figure 3.} (A) Transitions among the 4 distinct nucleosome states (i.e., $AR$, $UR$, $AU$, and $UU$) in the 4-state model. The time step is small enough that at most only one modification site of a nucleosome may change on each time step. (B) Transition probabilities between nucleosome states via recruitment conversions, where $X$ can either be $A$ or $U$ while $Y$ can either be $R$ or $U$. $k_{X}$ ($k_{Y}$)=2 if $X$ ($Y$) is $U$, otherwise $k_{X}$ ($k_{Y}$)$=1$. Thus, as an example, the transition probability  from the $AR$ state to the $AU$ state is the same as that from the $UR$ state to the $UU$ state. (C) Transition probabilities between nucleosome states  via exchange conversions, where $X$ can either be $A$ or $U$ while $Y$ can either be $R$ or $U$.
}
\label{figure3}
\end{figure}

\begin{figure}[!ht]
\begin{center}
\includegraphics[width=6in]{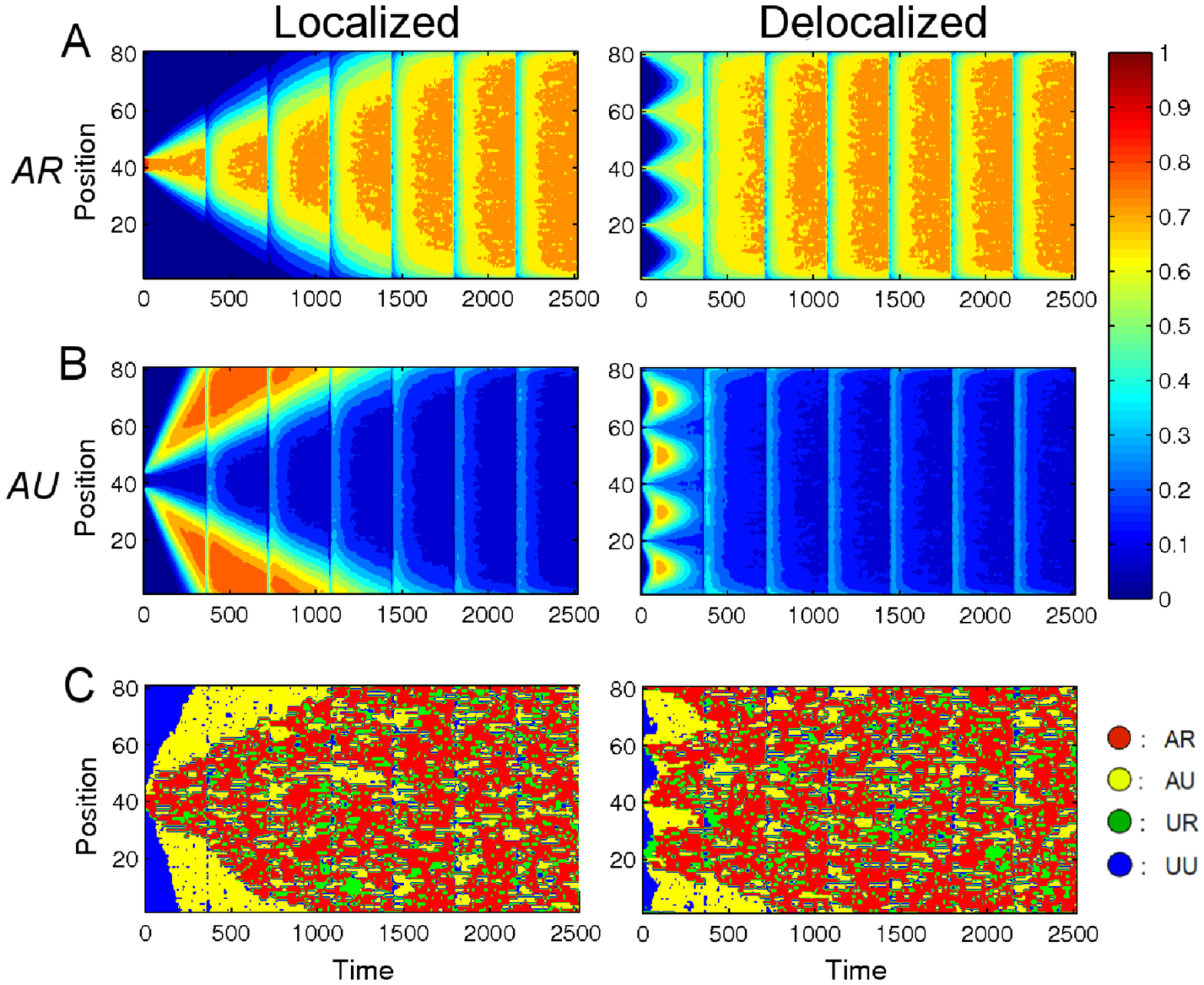}
\end{center}
\caption{
{\bf Figure 4.} Space-time plots of the average level of $AR$ and $AU$ nucleosomes for the \emph{localized} and  \emph{delocalized}  initial conditions are shown in (A) and (B).  Here by `level' we mean the fraction of runs for which the nucleosome is in the indicated state. These levels are computed by counting the indicated type of nucleosome in all runs at each position and time, and averaging over 2000 runs. The red color indicates a higher level of the indicated type of nucleosome while the blue color indicates a lower level of that type of nucleosome. (C) Space-time plots for a single run for the \emph{localized} and \emph{delocalized} initial conditions. $AR$, $AU$, $UR$, and $UU$ nucleosomes are plotted in red, yellow, green, and blue, respectively.
}
\label{figure4}
\end{figure}

\begin{figure}[!ht]
\begin{center}
\includegraphics[width=4in]{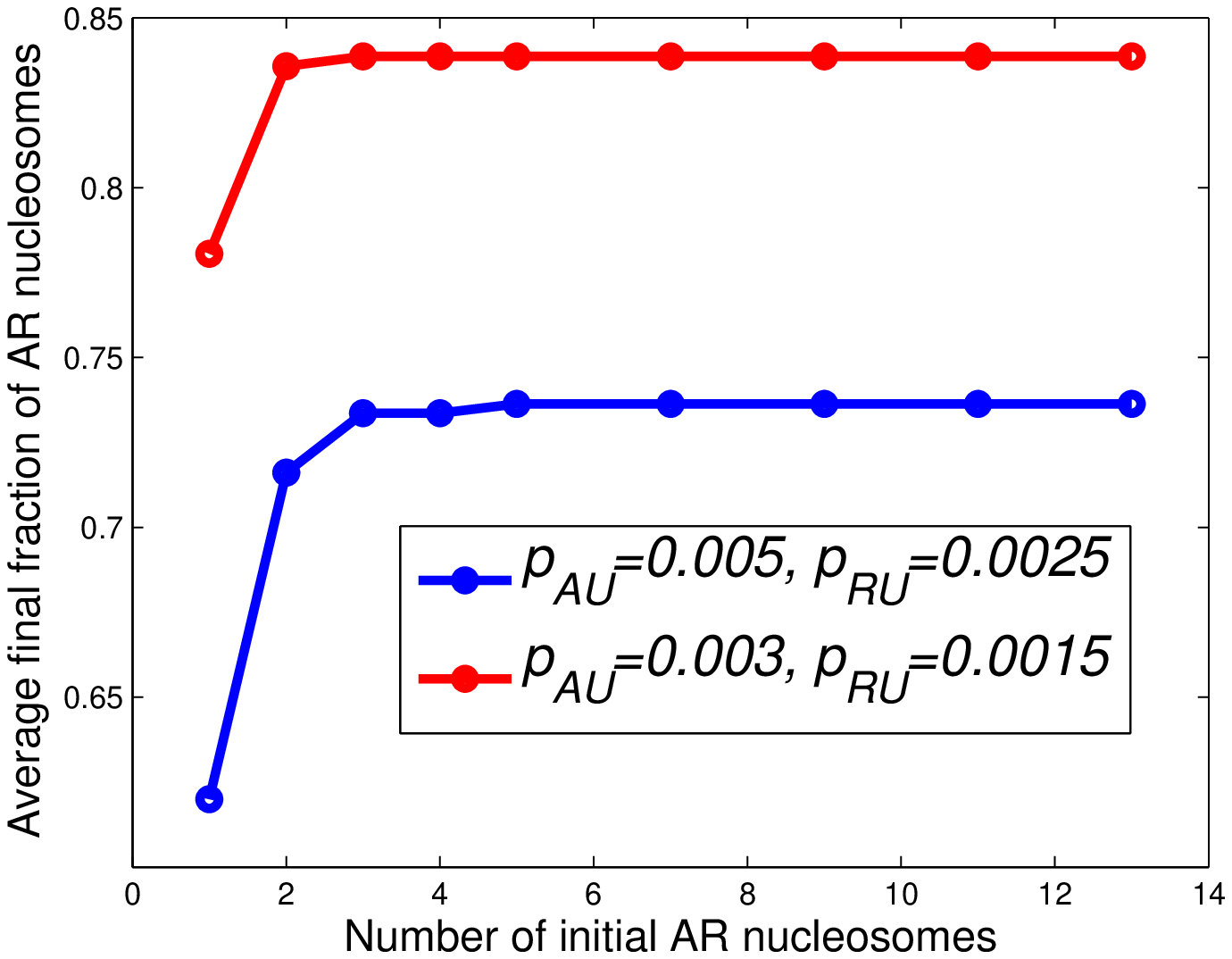}
\end{center}
\caption{
{\bf Figure 5.} The average final fraction of $AR$ nucleosomes is plotted as a function of $m$ (the number of initial $AR$ nucleosomes) for ($p_{_{AU}}$, $p_{_{RU}}$) being  (0.003, 0.0015)(red) and (0.005, 0.0025)(blue). These levels of $AR$ nucleosomes are computed by averaging the final number of $AR$ nucleosomes in the simulations over 2000 runs.
}
\label{figure5}
\end{figure}

\begin{figure}[!ht]
\begin{center}
\includegraphics[width=6in]{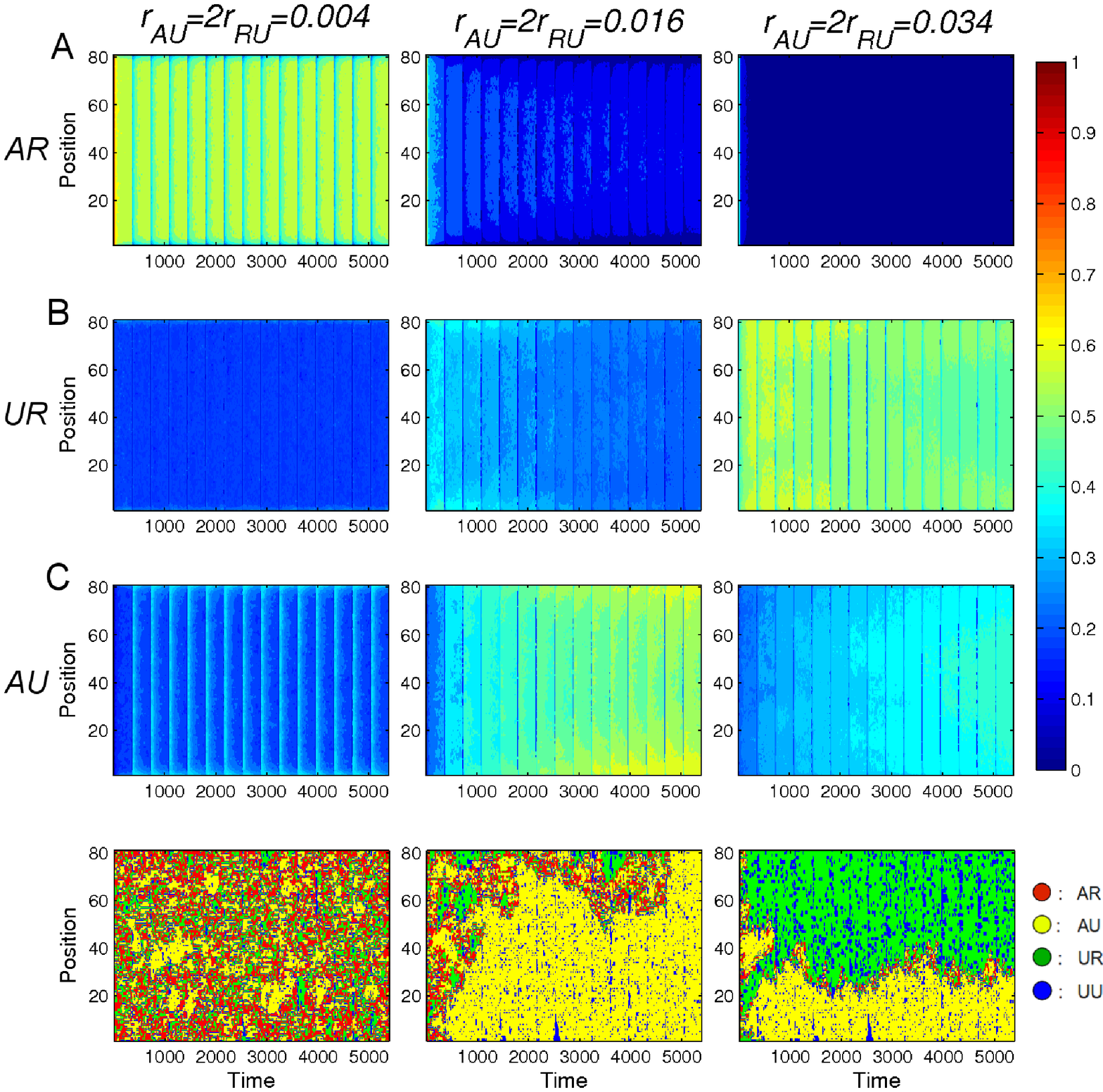}
\end{center}
\caption{
{\bf Figure 6.} In these plots, all nucleosomes are initially ($t=0$) in the $AR$ state. Space-time plots of the average level of $AR$, $UR$, and $AU$ nucleosomes for $r_{_{AU}}=0.004, 0.016$, and $0.034$ are shown in (A), (B), and (C), respectively. These plots are similar to Figure 4A and B. Here by ‘level’ we mean the fraction of runs for which the nucleosomes is the indicated state. (D) Space-time plots for a single run with $r_{_{AU}}=0.004$, $0.016$, $0.034$. $AR$, $AU$, $UR$, and $UU$ nucleosomes are plotted in red, yellow, green, and blue, respectively.
}
\label{figure6}
\end{figure}

\begin{figure}[!ht]
\begin{center}
\includegraphics[width=4in]{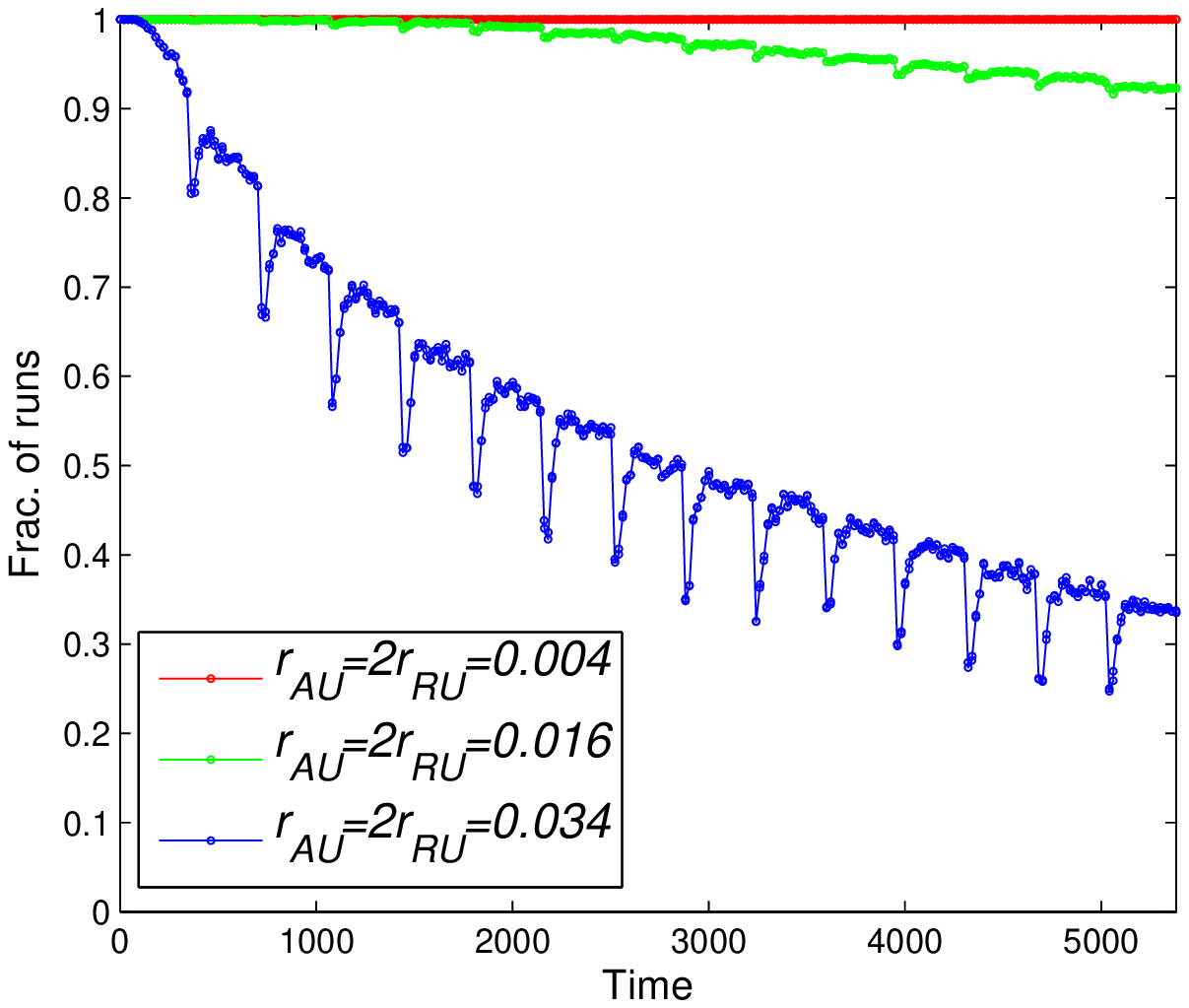}
\end{center}
\caption{
{\bf Figure 7.} The fraction of runs that have at least one $AR$ nucleosome on the lattice is plotted as a function of time for $r_{_{AU}}=0.004$, $0.016$, and $0.034$. 
}
\label{figure7}
\end{figure}

\begin{figure}[!ht]
\begin{center}
\includegraphics[width=4in]{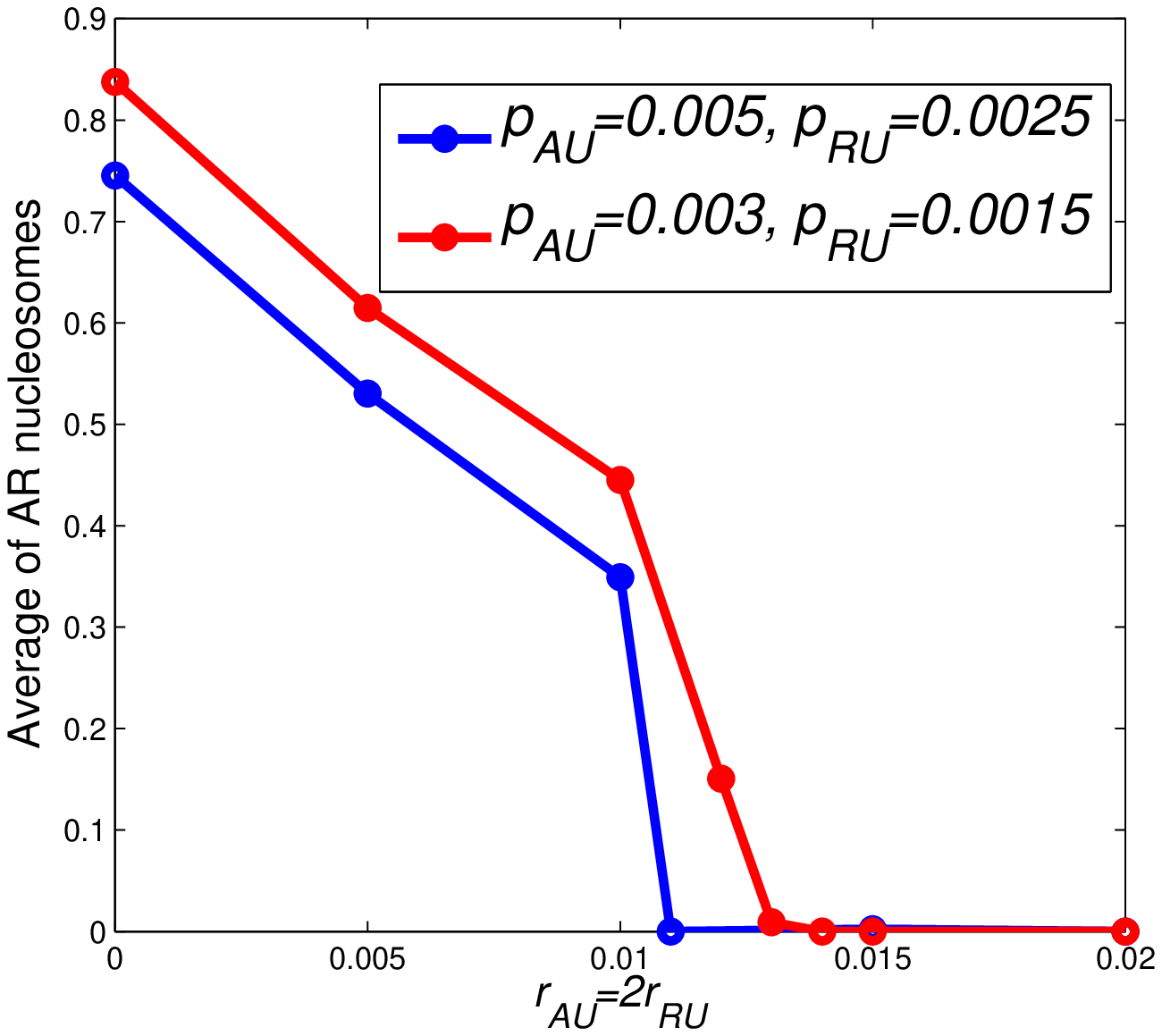} 
\end{center}
\caption{
{\bf Figure 8.} The average level of $AR$ nucleosomes is plotted as a function of $r_{_{AU}}$ for both $p_{_{AU}}=0.003$ and $0.005$. These levels of $AR$ nucleosomes are computed by averaging the final number of $AR$ nucleosomes in the simulations over all the runs and the whole lattice.
}
\label{figure8}
\end{figure}

\begin{figure}[!ht]
\begin{center}
\includegraphics[width=6in]{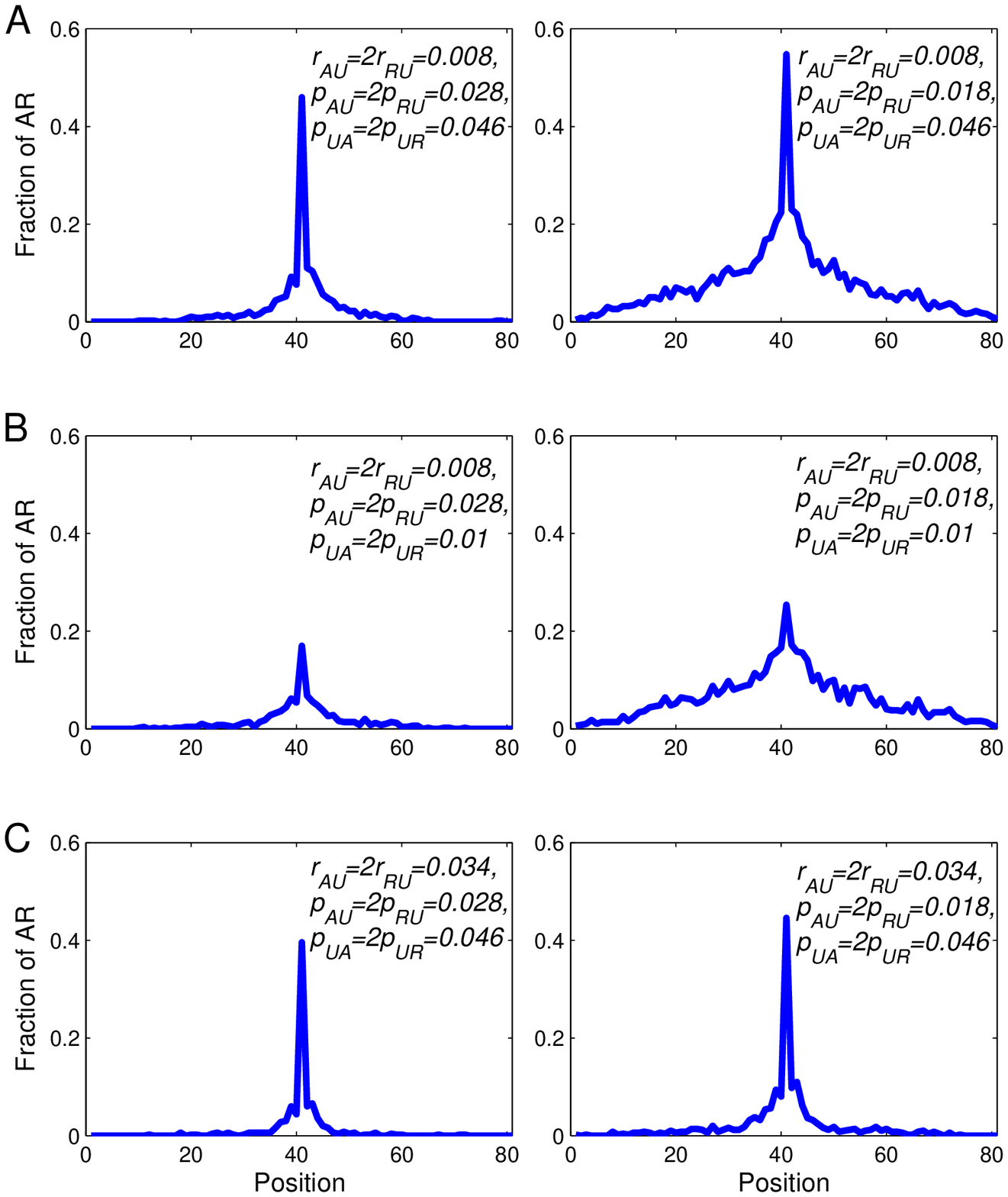} 
\end{center}
\caption{
{\bf Figure 9.}  Distributions of $AR$ nucleosomes are plotted at the final time of the simulations (time = 4000).  
}
\label{figure9}
\end{figure}

\begin{figure}[!ht]
\begin{center}
\includegraphics[width=4in]{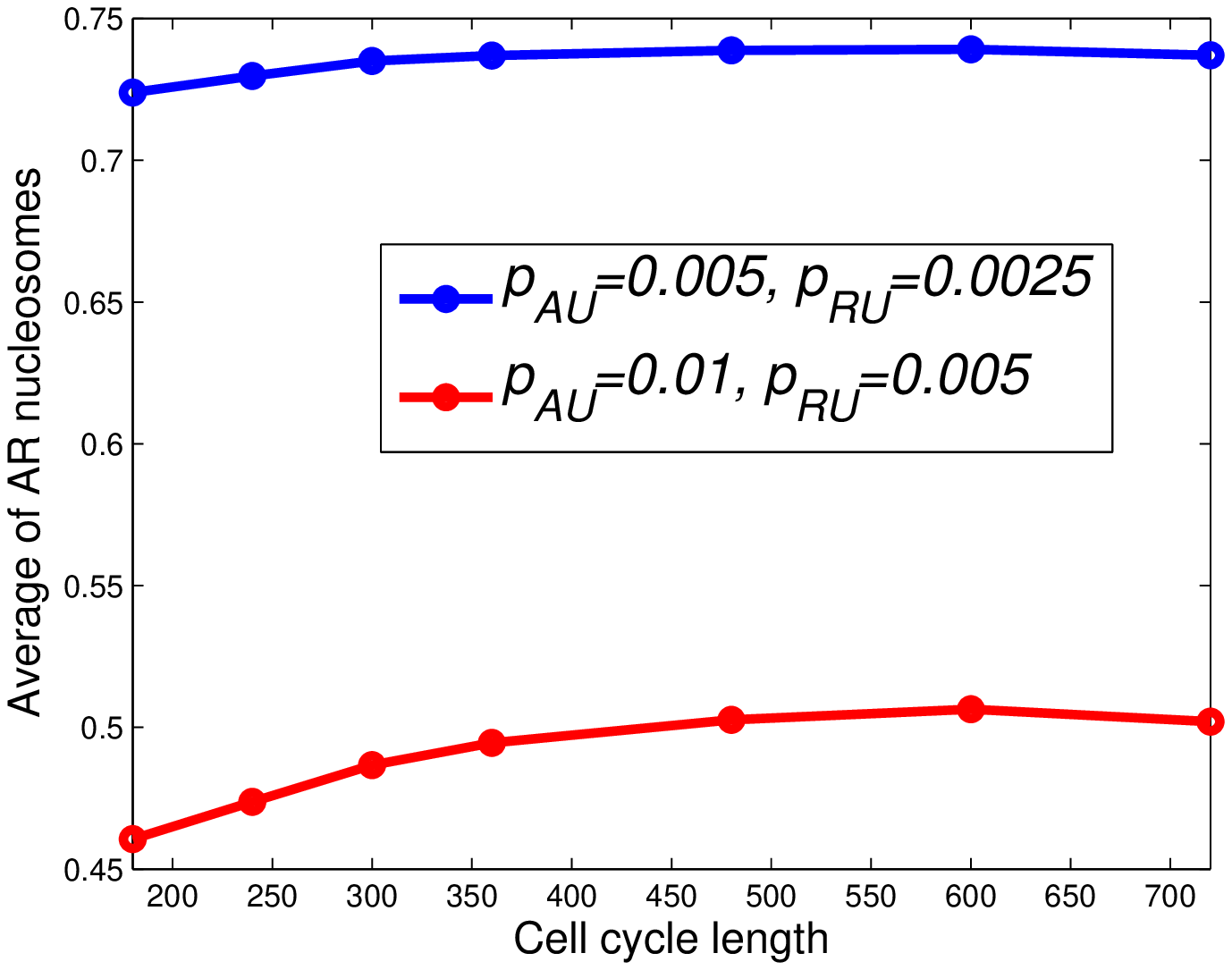}
\end{center}
\caption{
{\bf Figure 10.} The average $AR$ nucleosome level is plotted as a function of cell-cycle length. These levels of $AR$ nucleosomes are computed by averaging the final number of $AR$ nucleosomes in the simulations over all the runs and the whole lattice. $r_{_{UA}}=2r_{_{UR}}=0.046$, $p_{_{UA}}=p_{_{UR}}=0$, and $r_{_{AU}}=r_{_{RU}}=0$.
}
\label{figure10}
\end{figure}

\begin{table}
\caption{\textbf{Table 1 : Summary of parameters}}
\begin{center}
	\begin{tabular}{|p{2cm}|p{7cm}|p{7cm}|}
	\hline
		Parameters & Physical description & Biological process simulated \\\hline
		$r^{i}_{UR}$, $r^{i}_{UA}$ & Coefficient determining the probabilty of $U$ converting to $R$/$A$ via recruitment by the surrounding $R$/$A$ marks  & \textbf{Histone methylation spreading}: existing H3K27me3/H3K4me3 recruits methylase to methylate nearby nuclesomes.\\\hline
		$r^{i}_{RU}$, $r^{i}_{AU}$ & Coefficient determining the probabilty of $R$/$A$ converting to $U$ via recruitment by the surrounding $A$/$R$ marks & \textbf{Crosstalk between $A$ and $R$}: existing H3K27me3/H3K4me3 recruits demethylase to demethylate nearby H3K4me3/H3K27me3. \\\hline
		$p^{i}_{UR}$, $p^{i}_{UA}$ & Probabilty of $U$ converting to $R$/$A$ independent of the states of other nearby nucleosomes & \textbf{Nucleation :} continuous random histone marks placements at nucleosome site $i$ \\\hline
		$p^{i}_{RU}$, $p^{i}_{AU}$ & Probabilty of $R$/$A$ converting to $U$ independent of the states of other nearby nucleosomes  & \textbf{Histone turnover rate} : histone marks can also be lost by random demethylation. \\\hline
		$f^{i}_{R}$, $f^{i}_{A}$ & Fraction of $R$/$A$ marks in nucleosomes within the recruitment range $l$ of site $i$ & We assume that the probability of recruitment (involved in the methylation spreading and crosstalk processes above) is proportional to the local density of the recruiting mark.\\\hline
		$\tau$ & The cell-cycle DNA replication period &  \textbf{Cell cycle}\\\hline
		$l$ & The nucleosome interaction distance &  \textbf{Recruitment Range}\\\hline
	\end{tabular}
	\label{tab:}
\end{center}
\end{table}

\begin{table}[H]
\caption{\bf{Table 2: }Model paramters}
\begin{center}
\scalebox{0.8}{
    \begin{tabular}{ | l | c  | c | c |}
    \hline
    Dynamical processes & Parameters & Characteristic time & References \\ \hline
    Adding H3K4me3 marks via recruitment & $r_{_{UA}}$ & 0.5 -6 hours & \cite{ZeeBarry,Zee29012010}\\ \hline
	Adding H3K27me3 marks via recuritment & $r_{_{UR}} $& 0.5- 6 hours & \cite{ZeeBarry,Zee29012010} \\ \hline
    Removing both H3K4me3 and H3K27me3 marks via exchange & $p_{_{AU}}$ and $p_{_{RU}}$ & 1-24 hours & \cite{ZeeBarry,Zee29012010}\\ \hline
	Adding both H3K4me3 and H3K27me3 marks via exchange & $p_{_{UA}}$ and $p_{_{UR}} $& not known &$\hbox{--------}$\\ \hline
	Removing both H3K4me3 and H3K27me3 marks via Recruitment & $r_{_{AU}}$ and $r_{_{RU}}$ & not known &$\hbox{--------}$\\ \hline
	Cell cycle length in human embryonic stem cells  & $\tau$ & 12 hours & \cite{Orford2008115}\\\hline
	Cell cycle length in human adult cells & $\tau$& 24 hours &\cite{Orford2008115}\\ \hline
    \end{tabular}
    }
\end{center}
\end{table}

\end{document}